\pdfoutput=1
\documentclass[11pt,twoside,a4paper,cmspaper,final,collab]{cms-tdr}
\def\svnVersion{ec3c20e}\def\svnDate{2025/08/14}\def\cmsCernNoTag{CERN-EP-2024-326}\def\cmsCernDate{\today}\def\cmsMessage{Published in Physical Review Letters as \href{http://dx.doi.org/10.1103/srvm-f1h3}{\doi{10.1103/srvm-f1h3}.}}
\begin{document}\cmsNoteHeader{EXO-21-015}

\newcommand{\gq}{\ensuremath{g_{\PQq}}\xspace}
\newcommand{\kV}{\ensuremath{\kappa_{\PV}}\xspace}
\newcommand{\emu}{\ensuremath{\Pe\PGm}\xspace}
\newcommand{\mutauh}{\ensuremath{\PGm\tauh}\xspace}
\newcommand{\etauh}{\ensuremath{\Pe\tauh}\xspace}
\newcommand{\tauhtauh}{\ensuremath{\tauh\tauh}\xspace}
\newcommand{\tautau}{\ensuremath{\PGtp\PGtm}\xspace}
\newcommand{\WW}{\ensuremath{\PWp\PWm}\xspace}
\newcommand{\Zptautau}{\ensuremath{\zp\to\tautau}\xspace}
\newcommand{\ZpWW}{\ensuremath{\zp\to\WW}\xspace}
\newcommand{\mjj}{\ensuremath{m_{jj}}\xspace}
\newcommand{\mllmet}{\ensuremath{m(\Pell_{1},\Pell_{2},\ptmiss)}\xspace}
\newcommand{\memumet}{\ensuremath{m(\Pe,\PGm,\ptmiss)}\xspace}
\newcommand{\Wjets}{\ensuremath{\PW\!+\text{jets}}\xspace}
\newcommand{\Zjets}{\ensuremath{\PZ\!+\text{jets}}\xspace}
\newcommand{\Ztauhjets}{\ensuremath{\PZ(\to\!\tauhtauh)\!+\text{jets}}\xspace}
\newcommand{\PGff}{\HepParticle{f}{}{}\Xspace}
\newcommand{\PAGf}{\HepAntiParticle{f}{}{}\Xspace}
\newcommand{\gZff}{\ensuremath{g_{\PZ\PGff\PAGf}}\xspace}
\newcommand{\gZVV}{\ensuremath{g_{\zp\PW\PW}^{\text{max}}}\xspace}
\newcommand{\gZpVV}{\ensuremath{g_{\zp\PW\PW}}\xspace}
\newcommand{\gl}{\ensuremath{g_{1,\,2}}\xspace}
\newcommand{\Zmm}{\ensuremath{\PZ(\to\!\PGm\PGm)}\xspace}
\newcommand{\Wmn}{\ensuremath{\PW(\to\!\PGm\PGn)}\xspace}
\newcommand{\mZpr}{\ensuremath{m_{\zp}}\xspace}

\cmsNoteHeader{EXO-21-015}

\title{Search for a neutral gauge boson with nonuniversal fermion couplings in vector boson fusion processes in proton-proton collisions at \texorpdfstring{$\sqrt{s} = 13\TeV$}{sqrt(s) = 13 TeV}}

\date{\today}

\abstract{
The first search for a heavy neutral spin-1 gauge boson (\zp) with nonuniversal fermion couplings produced via vector boson fusion processes and decaying to tau leptons or \PW bosons is presented. The analysis is performed using LHC data at $\sqrt{s} = 13\TeV$, collected from 2016 to 2018 with the CMS experiment and corresponding to an integrated luminosity of 138\fbinv.  The data are consistent with the standard model predictions. Upper limits are set on the product of the cross section for production of the \zp boson and its branching fraction to $\PGt\PGt$ or $\PW\PW$. The presence of a \zp boson decaying to $\PGtp\PGtm$ ($\PWp\PWm$) is excluded for masses up to 2.45(1.60)\TeV, depending on the \zp boson coupling to standard model weak bosons, and assuming a \Zptautau ($\PWp\PWm$) branching fraction of 50\%.
}

\hypersetup{
pdfauthor={CMS Collaboration},
pdftitle={Search for a neutral gauge boson with nonuniversal fermion couplings in vector boson fusion processes in proton-proton collisions at sqrt(s) = 13 TeV},
pdfsubject={CMS},
pdfkeywords={CMS, heavy resonance, taus, BSM}}

\maketitle 

The standard model (SM) of particle physics has successfully explained many experimental observations. However, it is not an ultimate theory of nature. For example, the SM fails to explain neutrino masses and the matter-antimatter asymmetry of the Universe. Furthermore, recent experimental results from precision measurements, such as the ratios $R(\PDz)$ and $R(\PDst)$ from \PQB meson decays, show tension with the SM expectations~\cite{LHCb:2023rd,Altmannshofer_2021}. Various models of new physics have been proposed to address the shortcomings of the SM. Examples of such models include $U(1)$ extensions~\cite{Das:2021esm,Allanach:2024jls}, top-assisted technicolor models~\cite{Hill:1994hp}, Randall--Sundrum models~\cite{Randall:1999ee,Randall:1999vf, Davoudiasl_2001}, grand unified theories~\cite{Mohapatra:1985xm}, and $E(6)$ models~\cite{London:1986dk}. Such models predict a new heavy neutral gauge boson \zp that could be observed in proton-proton ($\Pp\Pp$) collisions at the CERN LHC.

Previous searches for \zp bosons have considered production via Drell--Yan processes with various \zp boson couplings \gq to light quarks~\cite{D0:1996qsx,D0:2010kuq,CDF:2011nxq,201282,201363,201757,Khachatryan_2017,ATLAS:2019erb,CMS:2021ctt,ATLAS:2017eiz,ATLAS:2020zms}. A widely used benchmark model in those searches is the sequential SM, which assumes a \zp boson with the same couplings to quarks and leptons as the SM \PZ boson.  In this model, 95\% confidence level bounds on the \zp boson mass \mZpr of up to 5\TeV have been established by the ATLAS~\cite{ATLAS:2019erb} and CMS~\cite{CMS:2021ctt} collaborations, assuming a \zp boson with a small decay width and considering decays to electron and muon pairs.  However, the bounds are significantly weaker under less stringent assumptions, \eg, ones that allow nonuniversal (i.e., generation-dependent) fermion couplings with small values of \gq, and can have large \zp boson decay widths~\cite{Faroughy:2016,Langacker:2008yv,Lynch:2000md}. These models could account for the tensions between SM predictions and precision measurements, and motivate a search for \zp resonances produced through vector boson fusion (VBF) processes, with the \zp boson decaying to a pair of \PGt leptons~\cite{Faroughy:2016,Florez:2016uob}.

This Letter presents the first search for a \zp boson with nonuniversal fermion couplings produced through VBF processes at the LHC, using 138\fbinv of data collected during 2016--2018 from $\Pp\Pp$ collisions at $\sqrt{s} = 13\TeV$.
In a VBF reaction, a quark from each of the colliding protons radiates an SM electroweak vector boson, and the merger of these produces a \zp boson.  The \zp boson can subsequently decay into a pair of \PGt leptons, which are accompanied by two jets from the scattered quarks. While the analysis is optimized for the \Zptautau final state, the possibility of a \zp boson decaying into a pair of \PW bosons motivates an alternate interpretation of the results. A representative Feynman diagram can be found in Ref.~\cite{Florez:2016uob}. Four \zp boson decay channels are utilized: \mutauh, \etauh, \tauhtauh, and \emu, where \tauh refers to a \PGt lepton that decays hadronically. In the \Zptautau decay mode, the electrons and muons are produced from the leptonic decays of \PGt leptons, while they may arise from $\PW\to\Pell\PGn$ decays in the case of the \ZpWW mode, where $\Pell = \Pe,\PGm,\PGt$.

The CMS apparatus~\cite{CMS:2008xjf} is a multipurpose, nearly hermetic detector, designed to trigger on~\cite{CMS:2020cmk,CMS:2016ngn,CMS:2024run2trigger} and identify electrons, muons, photons, and (charged and neutral) hadrons~\cite{CMS:2020uim,CMS:2018rym,CMS:2014pgm}. A global particle-flow (PF) algorithm~\cite{CMS:2017yfk} aims to reconstruct all individual particles (PF candidates) in an event, combining information provided by the all-silicon inner tracker and by the crystal electromagnetic (ECAL) and brass-scintillator hadron calorimeters, operating inside a 3.8\unit{T} superconducting solenoid, with data from the gas-ionization muon detectors embedded in the flux-return yoke outside the solenoid. 

The primary vertex is taken to be the vertex corresponding to the hardest scattering in the event, evaluated using tracking information alone, as described in Section 9.4.1 of Ref.~\cite{CMS-TDR-15-02}. Electron candidates are reconstructed by matching energy clusters in the ECAL with tracks in the inner tracker. A dedicated electron identification is used to distinguish electrons produced in hard scattering processes from charged hadrons and from electrons produced through photon conversions~\cite{CMS:2020uim}. Muons are reconstructed using the tracker and muon chambers and by requiring limited associated energy deposited in the calorimeters~\cite{CMS:2018rym}. The electron and muon selections impose an isolation requirement to suppress lepton candidates from misidentified jets, and genuine leptons from hadron decays. The isolation variable $I_{\Pell}$ is defined as the scalar \pt (transverse momentum) sum, divided by the lepton \pt, of charged and neutral PF candidates within a cone in $(\eta,\phi)$, with $\eta$ the pseudorapidity and $\phi$ the azimuthal angle. The radius of the cone is 0.4 (0.3) around the muon (electron) direction at the interaction vertex. The sum excludes the lepton under consideration as well as particles identified as arising from additional $\Pp\Pp$ interactions within the same or a nearby crossing of proton-beam bunches (pileup)~\cite{CMS-JME-18-001,PileupMitigation}. The isolation criterion is $I_{\Pell} < 0.15$~\cite{CMS:2018rym}. 

Jets are clustered using the anti-\kt clustering algorithm~\cite{Cacciari1, Cacciari2} with an angular distance parameter of 0.4. Identification criteria are applied to jet candidates to remove anomalous effects from the calorimeters and jets from secondary pileup interactions~\cite{CMS:2019ctu}. For jets with $\pt>30\GeV$ and $\abs{\eta} < 2.5$ $(> 2.5)$, the identification efficiency is ${>}98$ $(90)\%$ depending on pseudorapidity. The jet energy scale and resolution are corrected depending on the \pt and $\eta$ of the jet~\cite{CMS:2016lmd}.
The missing transverse momentum \ptvecmiss is the negative sum of the individual transverse momentum vectors of all PF candidates~\cite{CMS:2019ctu}. Its magnitude is \ptmiss.

Jets originating from the hadronization of \PQb quarks (\PQb jets) are tagged using algorithms that identify secondary vertices. The identification efficiency is 40\%--60\% for genuine \PQb jets, depending on \pt, $\eta$, the year of data collection, and the light-flavor quark or gluon jet misidentification rate is 0.1--0.9\%~\cite{BTV-16-002}. 

Each \tauh candidate is reconstructed using the hadrons-plus-strips algorithm~\cite{CMS:2018jrd} and is identified with an artificial neural network that combines variables related to isolation and the \PGt lepton lifetime to identify specific classes of \tauh decay modes~\cite{CMS:2022prd}. The tight-isolation working point is used to define the signal region, which results in a \tauh identification efficiency of 60\% with a 0.05--0.3\% probability for a jet to be misidentified as a \tauh, depending on the \pt and $\eta$ of the \tauh candidate~\cite{CMS:2022prd}. The loose-isolation working point, which is used for background estimation studies, has a \tauh identification efficiency of 80\% and a 0.1--1\% jet misidentification probability. To suppress muons and electrons misidentified as \tauh, a discriminator based on an artificial neural network is used that requires that the lead track of the \tauh candidate not be associated with a global muon signature or an electron bremsstrahlung shower~\cite{Pioppi:2009gjl}. The misidentification rate for muons (electrons) is 0.02\% (3.60\%) for a genuine \tauh identification efficiency of 80\%.

Data for this search were collected using single-lepton triggers for final states with muons or electrons, and a double-\tauh trigger for the $\tauh \tauh$ analysis~\cite{CMS:2016ngn}. The single muon (electron) trigger requires an isolated muon (electron) with a minimum \pt of 24 or 27\GeV (32 or 35\GeV), depending on the data-taking year. The double-\tauh trigger uses isolation criteria and $\pt(\tauh)$ thresholds of 32, 35, or 40\GeV, depending on the year. 

Events satisfying the trigger selections must pass additional offline lepton requirements targeting the $\zp\to\tautau$ and $\zp\to\WW$ decay chains. An event must contain exactly one pair of leptons having opposite-sign electric charge (OS). The muon in the \mutauh (\emu) channel is required to have $\pt>35\, (30)\GeV$ within $\abs{\eta} < 2.1$. The electron in the \etauh (\emu) channel is required to have $\pt > 55\, (10)\GeV$ within $\abs{\eta} < 2.1$. The \tauh candidates in the \tauhtauh (\mutauh and \etauh) channel are required to have $\pt>70\, (20)\GeV$ within $\abs{\eta} < 2.1$, where the trigger is fully efficient.  These requirements constitute the dilepton selection.

To reduce background from SM processes containing top quarks, events must contain no identified \PQb jet with $\pt > 30\GeV$ and $\abs{\eta} < 2.4$. The associated neutrinos from the \PGt lepton and \PW boson decays in the signal events generate sizable \ptmiss. We require $\ptmiss > 30\GeV$ to suppress the contribution from pure quantum chromodynamic processes (QCD multijet events).

The distinctive signature of events from VBF processes is the presence of two energetic jets with a large $\eta$ separation ($\Delta \eta$), located in opposite hemispheres of the detector. Therefore, in addition to the lepton pair, we require that each event contain two well-identified jets ($j_{1}$ and $j_{2}$), each with $\pt > 30\GeV$ and $\abs{\eta} < 5$, with a reconstructed dijet mass (\mjj) greater than 500\GeV, $\abs{\Delta\eta(jj)} > 4.2$, and $\eta_{j_{1}} \eta_{j_{2}} < 0$. These jets and the leptons must all be well separated from each other, satisfying the requirement $\DR\equiv\sqrt{\smash[b]{(\Delta\eta)^2+(\Delta\phi)^2}} > 0.4$. We refer to these requirements as the VBF criteria.  They are effective at suppressing the background by a factor of $10^2$--$10^4$, depending on the background process, while retaining 70\%--75\% of signal events, depending on \mZpr. We note that approximately 1\% of background events involve a reconstructed VBF dijet pair that includes a pileup jet, while such occurrences are exceedingly rare in signal events. The signal regions (SRs) comprise events that satisfy the dilepton selection criteria, the \PQb jet veto requirement, the \ptmiss requirement, and the VBF selection criteria.

As an approximation to the \zp boson mass, we form the invariant mass of the system comprising the visible lepton pair and the missing momentum,
\ifthenelse{\boolean{cms@external}}
{
\begin{equation*}
  \begin{aligned}
    m(\Pell_{1},&\Pell_{2},\ptmiss) = \\
    &\sqrt{(E_{\Pell_{1}} + E_{\Pell_{2}} + \ptmiss)^{2} - ( \vec{p}_{\Pell_{1}} + \vec{p}_{\Pell_{2}} + \ptvecmiss )^{2}},
  \end{aligned}
\end{equation*}  
}
{
\begin{equation*}
  \mllmet = \sqrt{(E_{\Pell_{1}} + E_{\Pell_{2}} + \ptmiss)^{2} - ( \vec{p}_{\Pell_{1}} + \vec{p}_{\Pell_{2}} + \ptvecmiss )^{2}},
\end{equation*}  
}
which serves as the main observable used to distinguish between signal and background.  The strategy is to search for a broad enhancement in the spectrum at large values of \mllmet. 

The dominant SM background processes contributing to the search are \PW and \PZ boson production (\Wjets and \Zjets), production of top quark-antiquark pairs (\ttbar), and QCD multijet production. The \Wjets and \Zjets events contain genuine lepton candidates, energetic jets, and \ptmiss from neutrinos. Background from \ttbar events is characterized by two \PQb jets in addition to genuine leptons. Events from the QCD multijet production process can be mistaken for signal if jets are misidentified as leptons. Smaller backgrounds include the production of events with a single \PQt quark and those with the production of two vector bosons.

Simulated samples of \Zjets, \Wjets, and single \PQt quark events are produced with the \MGvATNLO~2.6.0 program~\cite{Alwall_2014} at leading order (LO) in QCD. Events from \ttbar are generated with \POWHEG~2.0~\cite{Nason:2004rx, Frixione:2007vw, Alioli_2010} at next-to-leading order (NLO) precision. The LO \PYTHIA generator~\cite{Sjostrand:2014zea} is used to model
diboson processes.  The production of one or two vector bosons in
association with two jets exclusively through pure electroweak
interactions is simulated with LO \MGvATNLO 2.6.0, and the leading
electroweak QCD interference terms in the cross section are found to be
small~\cite{CMS-SMP-16-018}.  Pure electroweak $\Pp\Pp \to \zp jj$ signal samples are generated using \MGvATNLO at LO precision. We use the simplified model in Ref.~\cite{Florez:2016uob}, where the \zp boson has \PZ-like couplings, and the \zp boson masses and couplings to SM particles are free parameters. The \zp boson coupling to first- and second-generation fermions is defined as $\gl\gZff$, where \gZff is the SM \PZ boson coupling, and \gl is a modifier for the coupling. The coupling to tau leptons is similarly defined with $g_{3}$ as the modifier for the SM \PZ boson coupling to tau leptons. Finally, the \zp boson coupling to the SM \PW boson is defined as $\gZpVV = \kV\gZVV$, where \gZVV is the maximal \zp boson coupling to the SM \PW boson dictated by the unitarity constraint~\cite{Florez:2016uob}, and \kV is a modifier for that coupling. The parameter \kV affects both the signal cross section and event kinematic properties. We do not consider anomalous couplings between the \zp boson and the \PZ/$\gamma$ bosons of the SM, and thus \zp decays to \PZ/$\gamma$ are not considered. Four sets of signal models are utilized: ($i$) \Zptautau decays with $\gl = 0$, henceforth referred to as simplified phenomenological model~1 (SPM1); ($ii$) \Zptautau decays with $\gl = 1$ (SPM2); ($iii$) $\zp \to \PW\PW$ decays with $\gl = 0$ (SPM3); and ($iv$) $\zp \to \PW\PW$ decays with $\gl = 1$ (SPM4). The $\gl = 0$ cases are a proxy for nonuniversal fermion coupling scenarios, where the couplings of the \zp boson to light fermions are suppressed. For all sets of signal models described above, five \kV values are considered, $\kV = \{ 0.1, 0.25, 0.50, 0.75, 1.0 \}$, for \zp masses ranging from 250 to 2500\GeV in steps of 250\GeV. The parameter $g_{3}$ is also varied to study the effect of different \Zptautau branching fractions, and results are presented as a function of the branching fractions.

The \MGvATNLO and \POWHEG generators used to simulate signal and background processes are interfaced with \PYTHIA~8.212 (8.230) using the CUETP8M1 (CP5) tune~\cite{Khachatryan:2015pea,Sirunyan:2019dfx} for parton shower and fragmentation in the 2016 (2017--2018) simulated samples, respectively. The NNPDF3.0 (3.1) NNLO~\cite{Ball:2014uwa,NNPDF:2017mvq} parton distribution functions (PDFs) are used in the event generation for 2016 (2017--2018) samples. The simulated background yields are normalized to the integrated luminosity using next-to-NLO (NNLO) or NLO cross sections  in QCD, while signal production cross sections are calculated at LO precision. The response of the CMS detector in these Monte Carlo simulated samples is simulated using dedicated software based on the \GEANTfour toolkit~\cite{Agostinelli:2002hh}. Pileup is incorporated by simulating additional interactions that are both in time and out of time with the collision.

For the \mutauh and \etauh channels, the main background comes from events with jets misidentified as \tauh candidates, mainly from processes such as \Wjets and \ttbar with one leptonic \PW boson decay. This background contribution is estimated from data using control regions (CRs) obtained with no VBF criteria and containing a \tauh candidate that satisfies the loose, but fails the tight-isolation requirement [antiISO \tauh]. Events in this CR are weighted with a transfer factor given by the product of a VBF selection efficiency, and a misidentification ratio [misID] defined as the ratio of the number of events with a nominal \tauh candidate to the number of those with an antiISO \tauh. The VBF selection efficiency is determined using a $\Wmn\!+\text{jets}$ CR with transverse mass $\mT(\vec{p}_{\PGm},\ptvecmiss)$ near the $m_{\PW}$ Jacobian peak, and is taken to be the fraction of events in this CR that also satisfy the VBF criteria. The misID ratios are derived as a function of $\pt(\tauh)$ and $\eta(\tauh)$ in a $\Zmm + \tauh$ CR.  Nearly all \tauh candidates in this CR arise from misidentified jets. The antiISO \tauh requirement used to determine the misID ratio can slightly bias the angular separation between the \tauh candidate and \ptvecmiss, and as a result can slightly bias the \mllmet shape. Therefore, any residual \mllmet shape biases in the misID \tauh background prediction are accounted for using correction factors as a function of $\Delta\phi(\vec{p}_{\Pell},\ptvecmiss)$, derived from the Monte Carlo simulated samples, and validated with simulation and data.

The dominant genuine-\tauh background contribution in the \mutauh and \etauh signal regions arises from \ttbar fully leptonic events. This background is estimated from simulation scaled by data-to-simulation efficiency factors (SFs) to account for mismodeling of the selection efficiencies in simulation. The SFs are measured in a \ttbar-enriched CR requiring the same lepton pair and VBF criteria as for the SR, but with the \PQb jet veto replaced by the requirement of one \PQb jet.

For the \tauhtauh channel, the main background contribution comes from \Zjets processes with genuine \tauh candidates from  $\PZ\to\tauh\tauh$, and is estimated from simulation with SFs obtained from a \PZ-enriched CR. This CR is selected with the same lepton-pair and VBF criteria as those of the SR, but with $\ptmiss < 30\GeV$ and the reconstructed mass $m(\tauhtauh)$ of the two \tauh candidates between 50 and 100\GeV. An SF is measured from these events, which is used to normalize the \Zjets background yield in the SR. To disentangle the individual contributions of the VBF and non-VBF selections to the overall SF used in the background estimation, the same \PZ-enriched CR is also studied without the VBF criteria. The normalization SF does not depend on \mllmet. We correct the yields in these \Ztauhjets CRs for the contamination from QCD multijet events estimated from data in CRs with same-sign electric charge (SS) \tauh pairs.  The transfer factor between SS and OS events is calculated using events with antiISO \tauh and $m(\tauhtauh) >100\GeV$.

The contribution of QCD multijet events in the \tauhtauh SR is determined using a QCD-enriched CR selected by the requirement of an SS \tauh pair. The transfer factor between this CR and the SR is obtained as the ratio of OS to SS \tauh pairs in CRs that fail the VBF and \ptmiss criteria and have an antiISO \tauh pair.

The smaller \Wjets background contribution in the \tauhtauh SR is determined by correcting the prediction from simulation with SFs derived from high-purity $\Wmn\!+\text{jets}$ control samples. 

In the \emu channel, the main background comes from \ttbar production with two leptonic \PQt quark decays. An estimation of the \ttbar background is performed by defining four regions in data. Region $A$ is the SR, and the events in CR $D$ differ from those in $A$ by failing the VBF criteria.  Regions $B$ and $C$ are defined by the same requirements as for $A$ and $D$ respectively, but with one rather than zero \PQb jets. The \ttbar yield $N_A$ in the SR for each \memumet bin is obtained as $N_A = (N_B/N_C)N_D$, where $N_B$ and $N_C$ are the data yields in those respective CRs, and $N_D$ is given by the yield in CR $D$ minus an estimate from simulation of the contribution of non-\ttbar processes. 

The internal consistency (closure) of all background estimation methods, in all channels, is validated in simulation, with additional tests performed in data CRs. These closure tests demonstrate that the background determination techniques reproduce the expected distributions in both rate and shape.

Various effects impact the shape and normalization of the predicted \mllmet background distribution. A leading source of systematic uncertainty is the difference in the misID \tauh ratios between light-quark and gluon jets.
We estimate this uncertainty by comparing the misID ratio measured in an SS $\Wmn+\tauh$ CR, which has a higher fraction of light-quark jets relative to gluon jets, with the nominal one measured in the $\Zmm+\tauh$ CR.  The difference is found to depend on $\pt(\tauh)$, varying from 4 to 20\%. The statistical uncertainties of the data in the CRs used for measuring SFs give rise to uncertainties in the background predictions of 1\%--21\%, depending on the process. Another source of systematic uncertainty ($<$20\% for all processes) is the closure of the background estimation methods.

The signal and background yields estimated from simulation are affected by similar sources of systematic uncertainty, with small differences between the 2016, 2017, and 2018 datasets. The integrated luminosities for the 2016, 2017, and 2018 data-taking years have 1.2\%--2.5\% individual uncertainties~\cite{CMS-LUM-17-003,CMS-PAS-LUM-17-004,CMS-PAS-LUM-18-002}, while the overall uncertainty for the 2016--2018 period is 1.6\%. Uncertainties associated with the identification of \PGm and \Pe are 1\%--2\% for all background and signal processes~\cite{CMS:2020uim,CMS:2018rym}. The uncertainty from the \tauh identification and isolation requirements is 6\%--9\%, depending on the data-taking year. The uncertainties in the identification and energy scales for \PGm, \Pe, and \tauh  are correlated across processes with genuine \PGm, \Pe, and \tauh candidates, and correlated across \mllmet bins. The uncertainty in the \tauh energy scale amounts to 2\% uncertainty in the yields. For the electrons misidentified as \tauh candidates, the identification uncertainties are 1\%--6.5\% depending on $\pt(\tauh)$. The jet energy scale uncertainties (2\%--5\% depending on $\eta$ and \pt) result in an uncertainty of 1\%--3\% in the signal yields, depending on \mllmet. Uncertainties due to the modeling of secondary pileup interactions are less than 1\%. The uncertainty in event acceptance from the PDF set used in simulation is determined according to the PDF4LHC recommendations~\cite{PDF4LHCRunII} by comparing the outputs from the CTEQ6.6L, MSTW08, and NNPDF10 PDF sets~\cite{Nadolsky:2008zw,mrst2006,nnpdf10} to those of the default PDF set.  This results in at most a 6\% uncertainty in the signal yield, and is correlated across data-taking years. The uncertainty in the \PQb tagging efficiency results in 2\%--9\% uncertainty in the predicted yields, depending on the process, and is only correlated between backgrounds with top quarks. The trigger uncertainty is 3\% for background and signal yields, and is uncorrelated across data-taking years. During the 2016 and 2017 data-taking periods, a gradual shift in the timing of the inputs of the ECAL L1 trigger in the region at $\abs{\eta} > 2.0$ caused a specific trigger inefficiency, for which correction factors and uncertainties are computed from data and applied to the acceptance derived from simulation, resulting in a 5\%--8\% uncertainty in the yields, depending on the process.
 
\begin{figure*}[tbh!]
\centering
\includegraphics[width=0.45\textwidth]{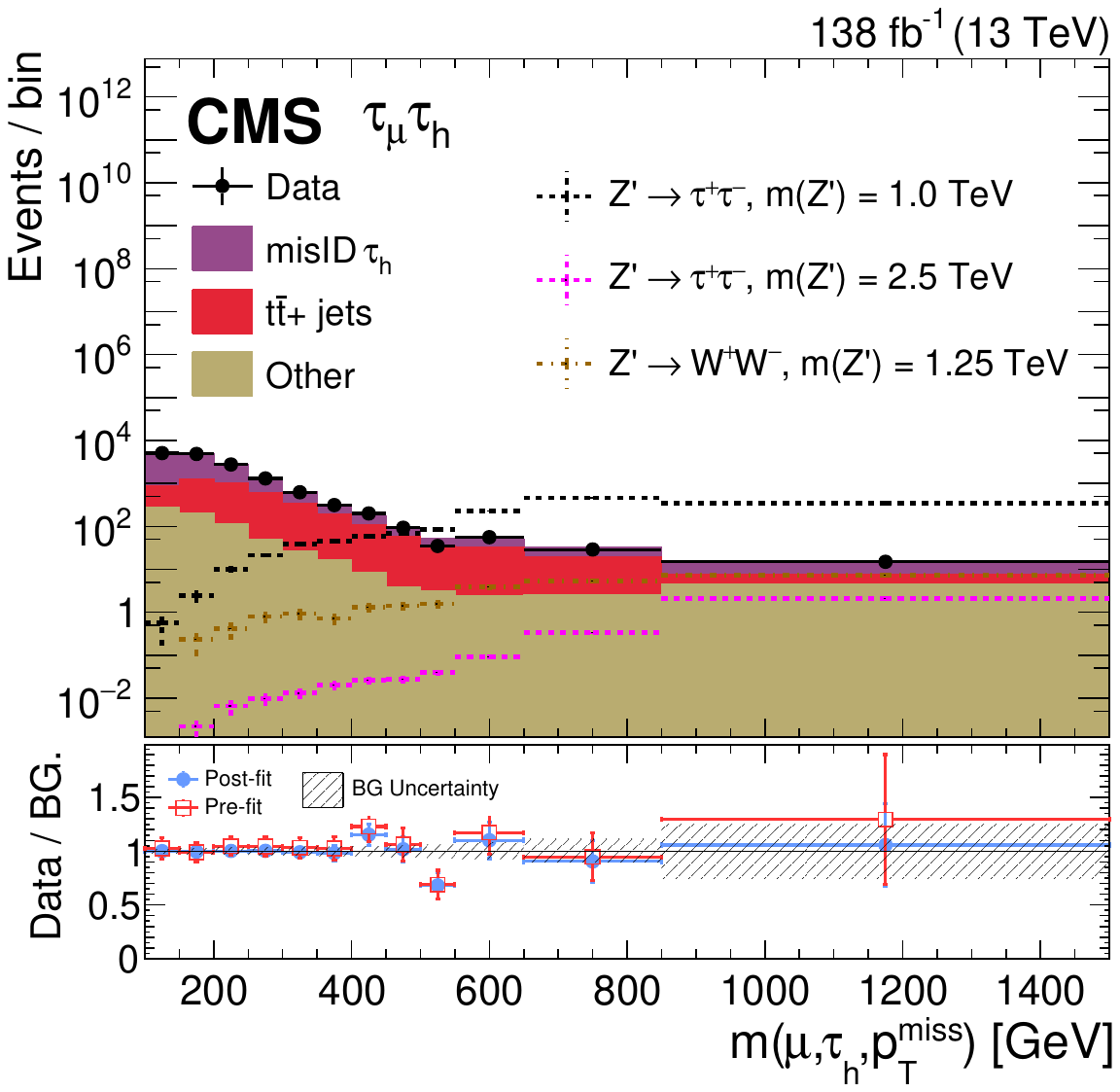}
\includegraphics[width=0.45\textwidth]{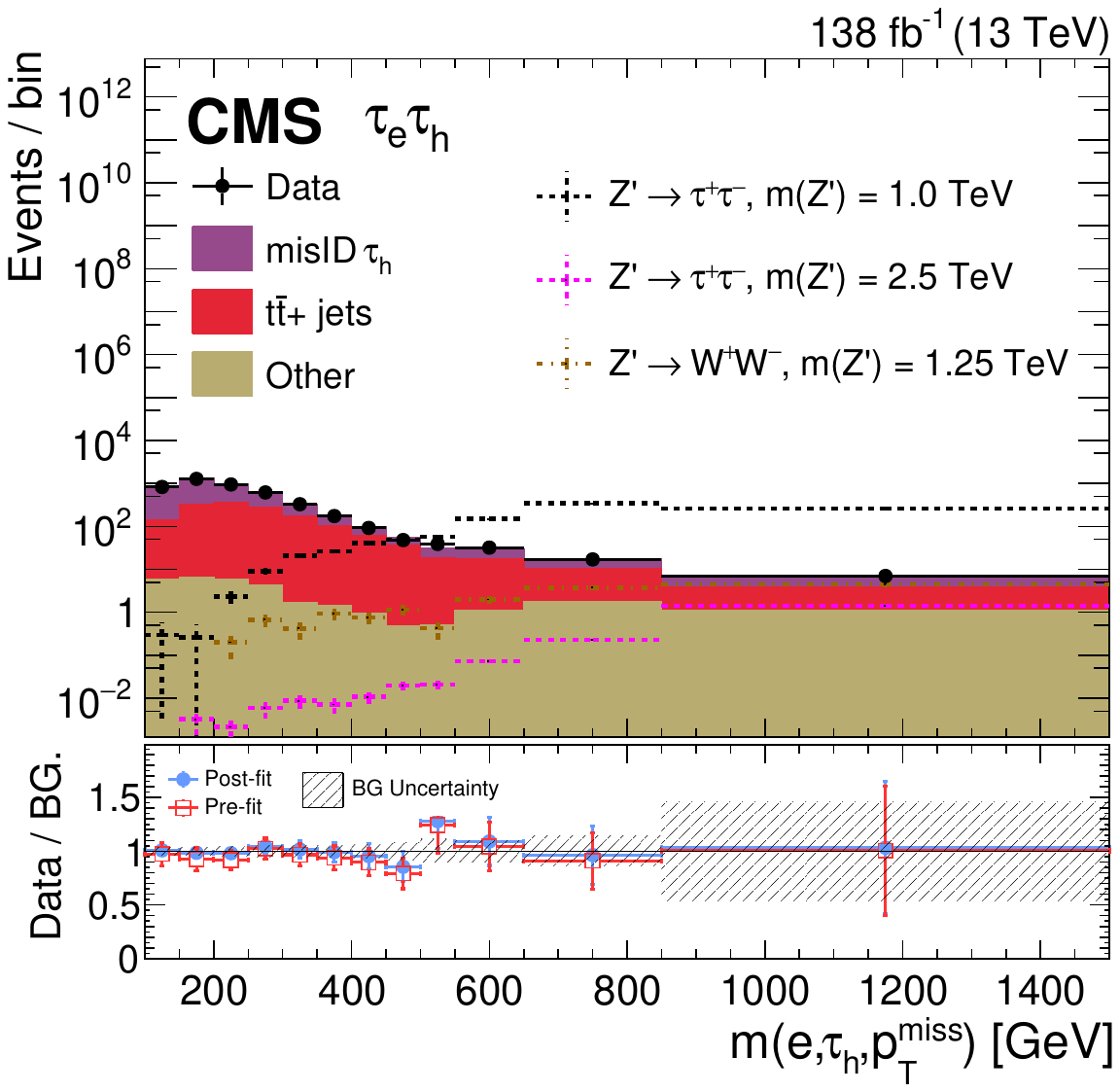}\\
\includegraphics[width=0.45\textwidth]{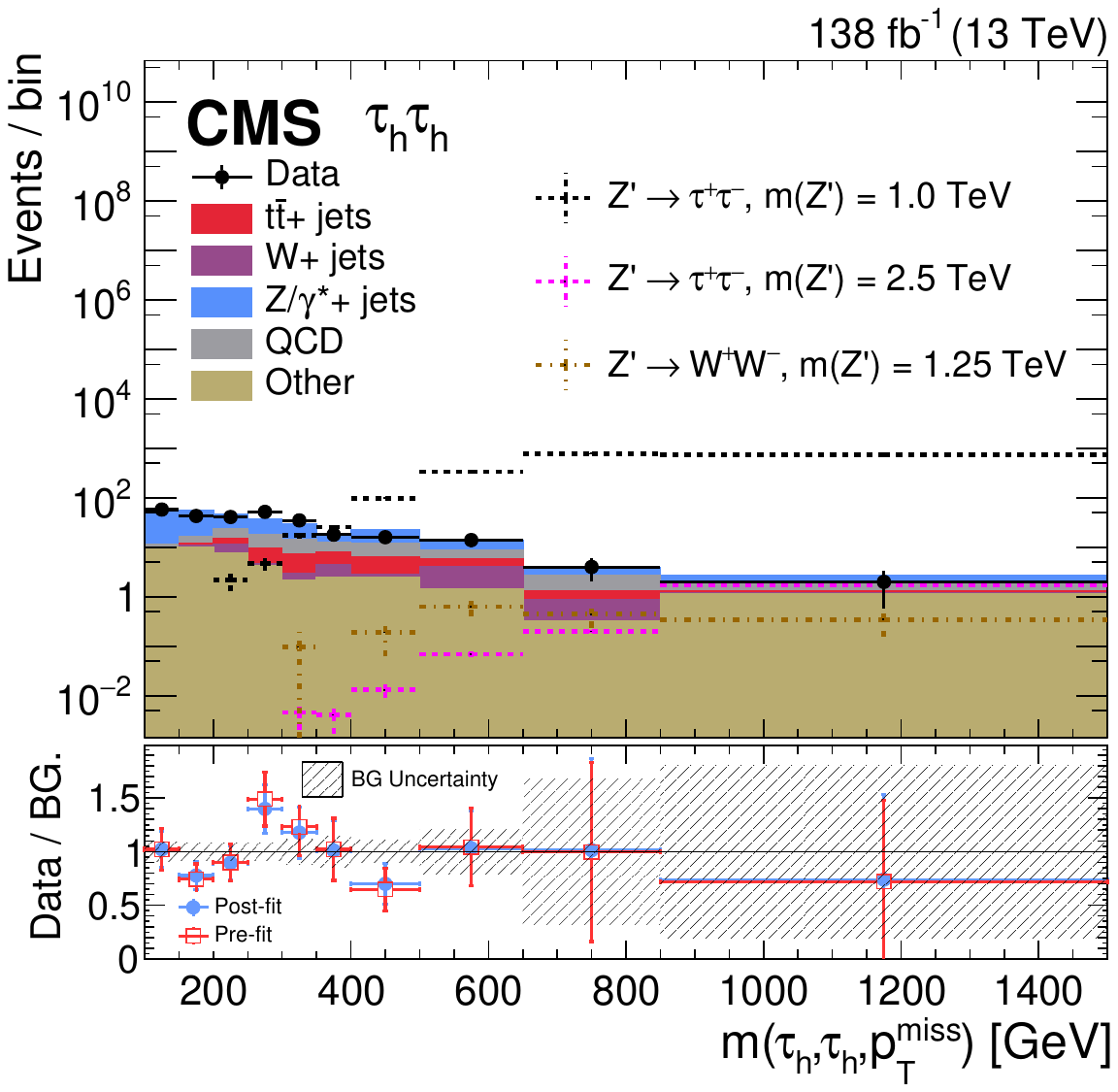}
\includegraphics[width=0.45\textwidth]{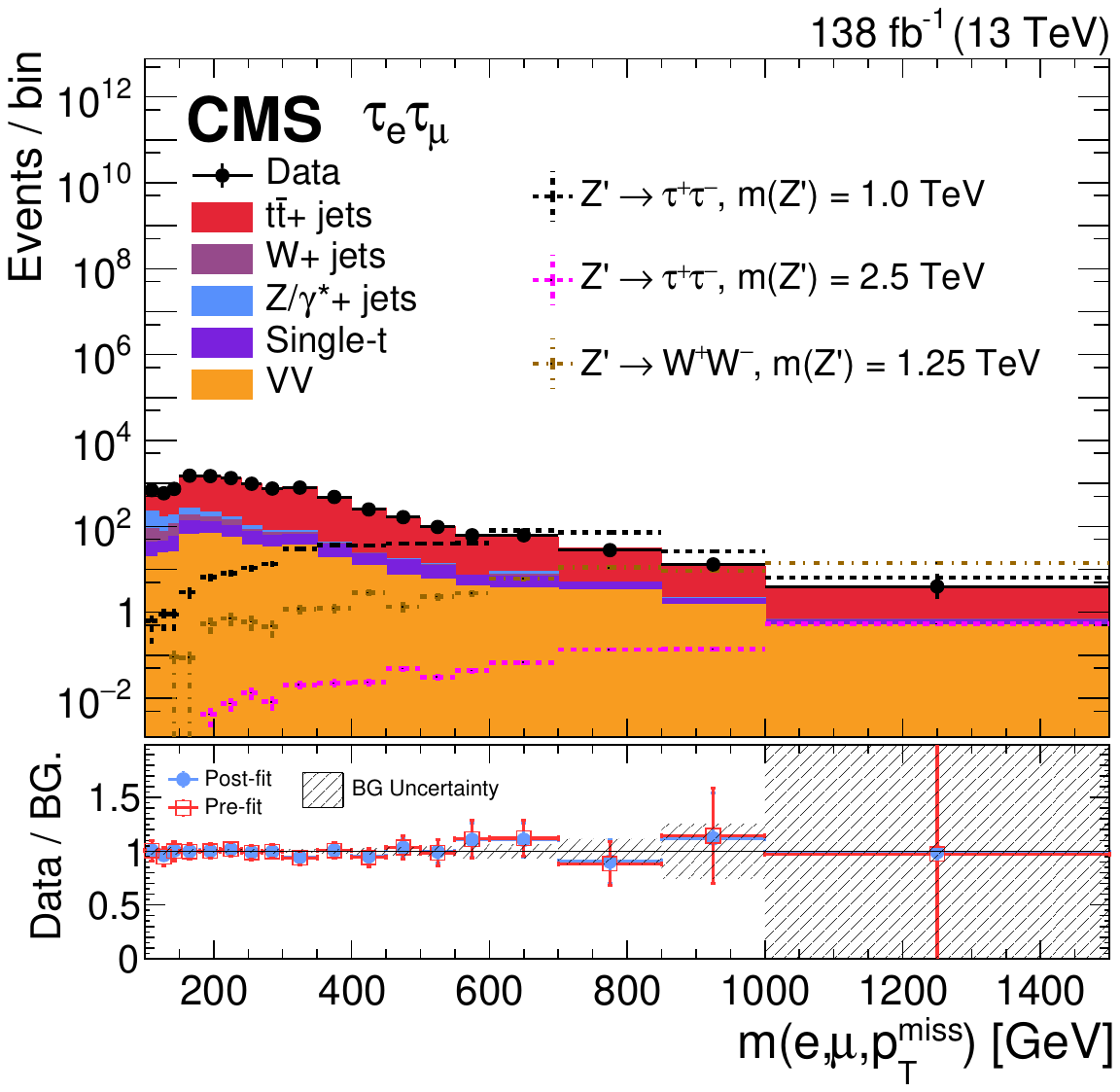}
\caption{Observed \mllmet for the data, and the postfit backgrounds (stacked histograms), in the signal region for the \mutauh (upper left), \etauh (upper right), \tauhtauh (lower left), and \emu (lower right) channels. The lower panels show ratios of the data to the prefit background prediction and postfit background yields as red open squares and blue points, respectively. The gray band in the lower panels indicates the systematic component of the postfit uncertainty. The dashed lines correspond to the signal expectation for \zp boson masses of 1\TeV (black) and 2.5~TeV (magenta) decaying to \tautau, normalized to cross sections of 199.4\unit{fb} and 0.7504\unit{fb} respectively. The dashed brown line corresponds to a \zp boson mass of 1.25\TeV decaying to \WW, normalized to a cross section of 61.14\unit{fb}. The signal cross sections correspond to $\{\kV,g_{3},\gl \} = \{ 1,1,0\}$.}
\label{fig:fakeratio}
\end{figure*}

Figure~\ref{fig:fakeratio} shows the \mllmet distribution for events in the SRs for all channels. The binning was chosen among several binning options as the one with the best signal significance and discovery reach in \mZpr. Backgrounds are estimated using either fully data-driven methods or simulation-based approaches, with CR-derived SFs applied to the SRs, as described above. The observed mass spectra shown in Fig.~\ref{fig:fakeratio} are consistent with the SM predictions. We set 95\% confidence level (\CL) upper limits on the product of the VBF \zp cross section and the branching fraction for the decay of the \zp boson to \tautau or \WW.  The limits are estimated following the modified frequentist construction \CLs~\cite{Cowan:2010js,Junk,CLs} using the CMS statistical analysis tool \textsc{Combine}~\cite{CAT-23-001}. Maximum likelihood fits are performed using the SR distributions to construct a combined profile likelihood ratio test statistic in bins of \mllmet. We use the asymptotic approximation to the profile likelihood test statistic. Systematic uncertainties are implemented as nuisance parameters, which are profiled and modeled with gamma or log-normal priors for the normalization parameters and Gaussian priors for the shape uncertainties. The upper limits on signal cross section are then used to derive exclusion bounds. 

\begin{figure*}[tbh!]
\centering
\includegraphics[width=0.49\textwidth]{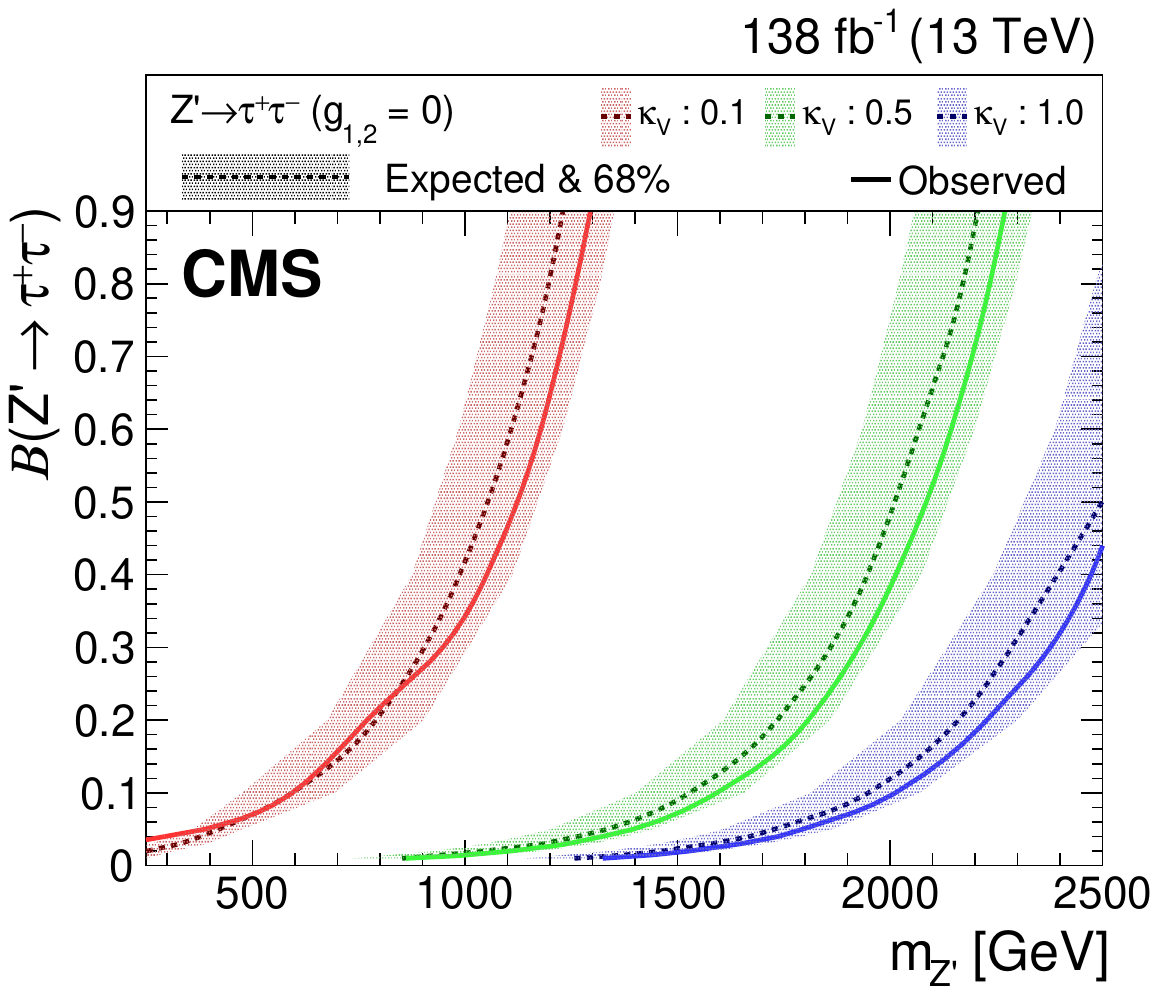}
\includegraphics[width=0.49\textwidth]{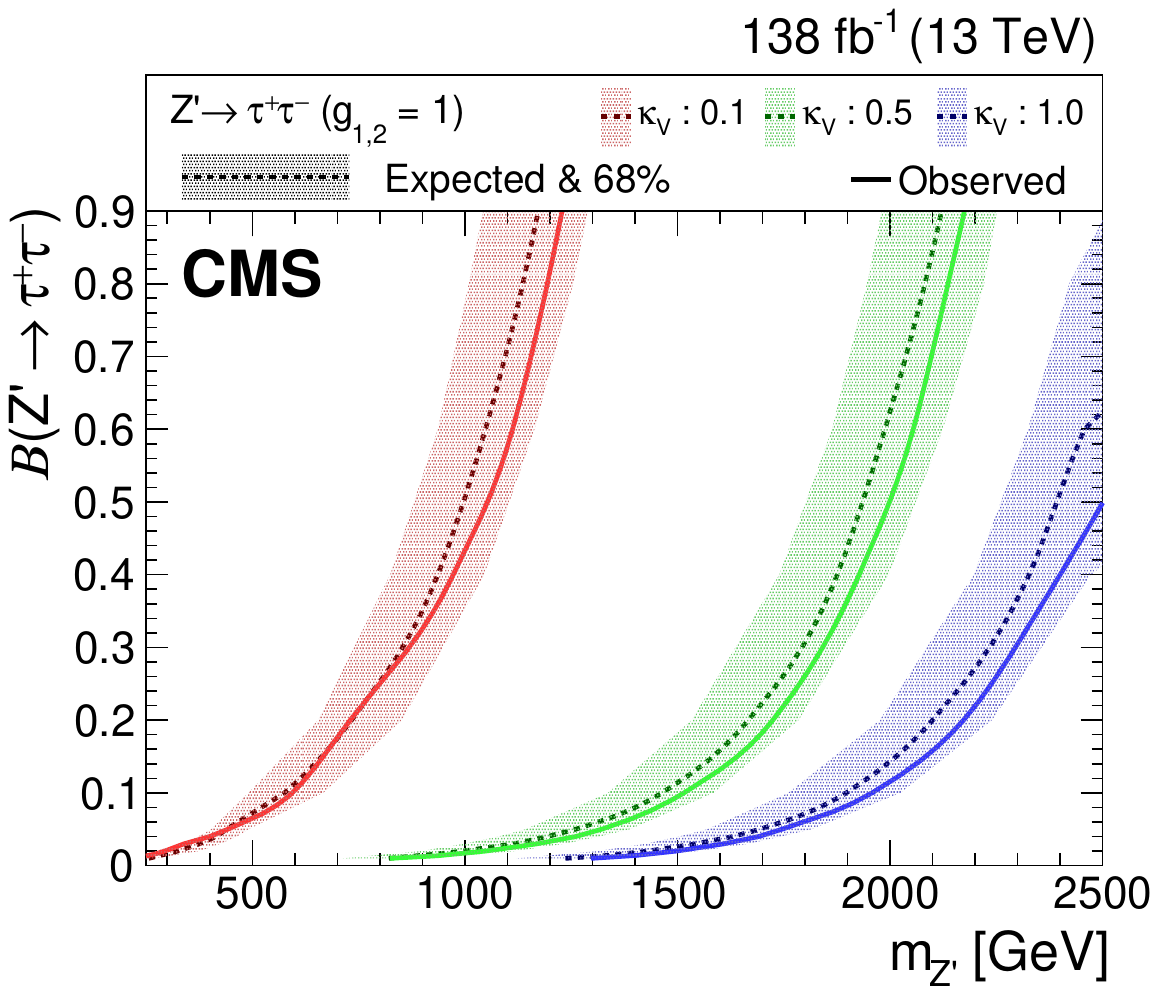}\\
\includegraphics[width=0.49\textwidth]{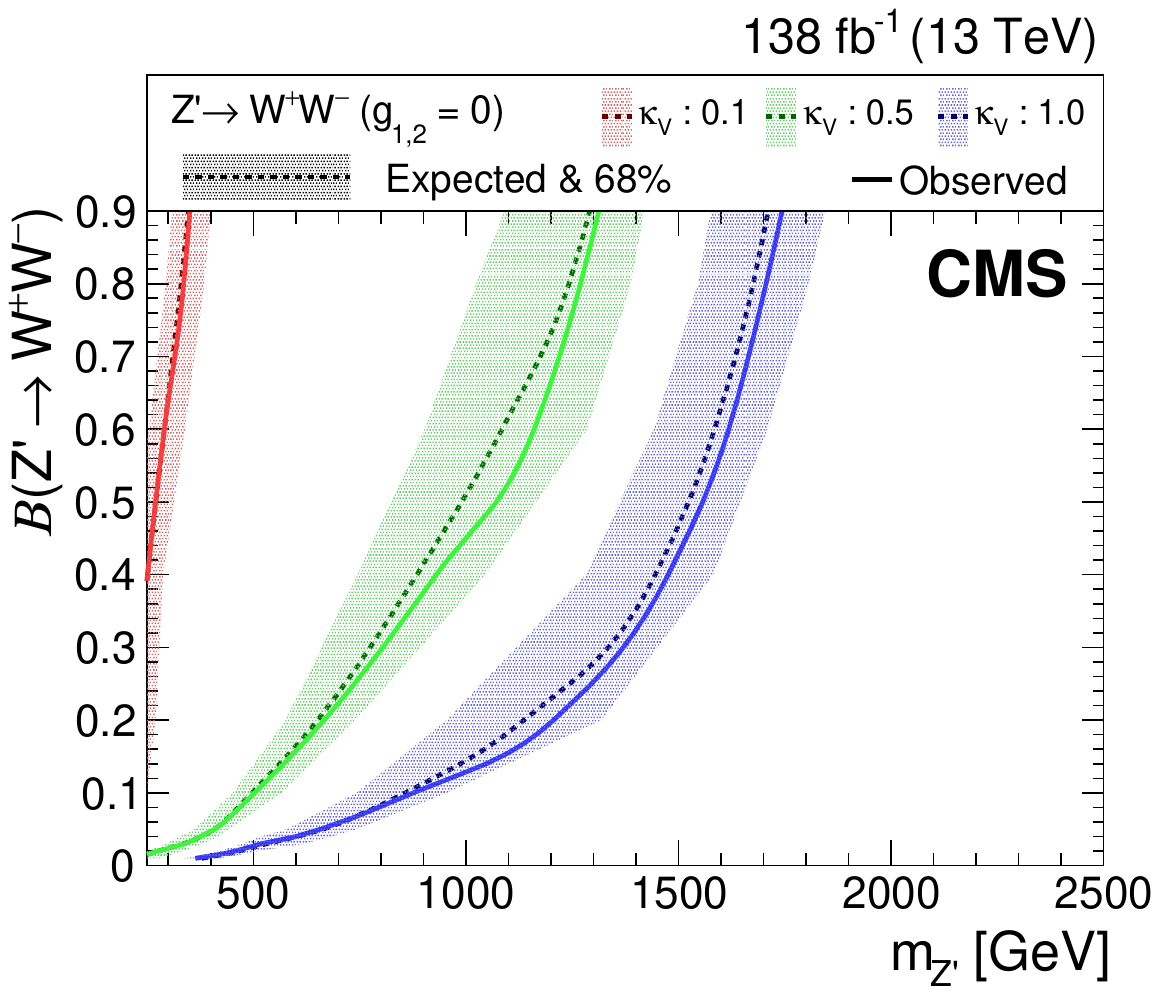}
\includegraphics[width=0.49\textwidth]{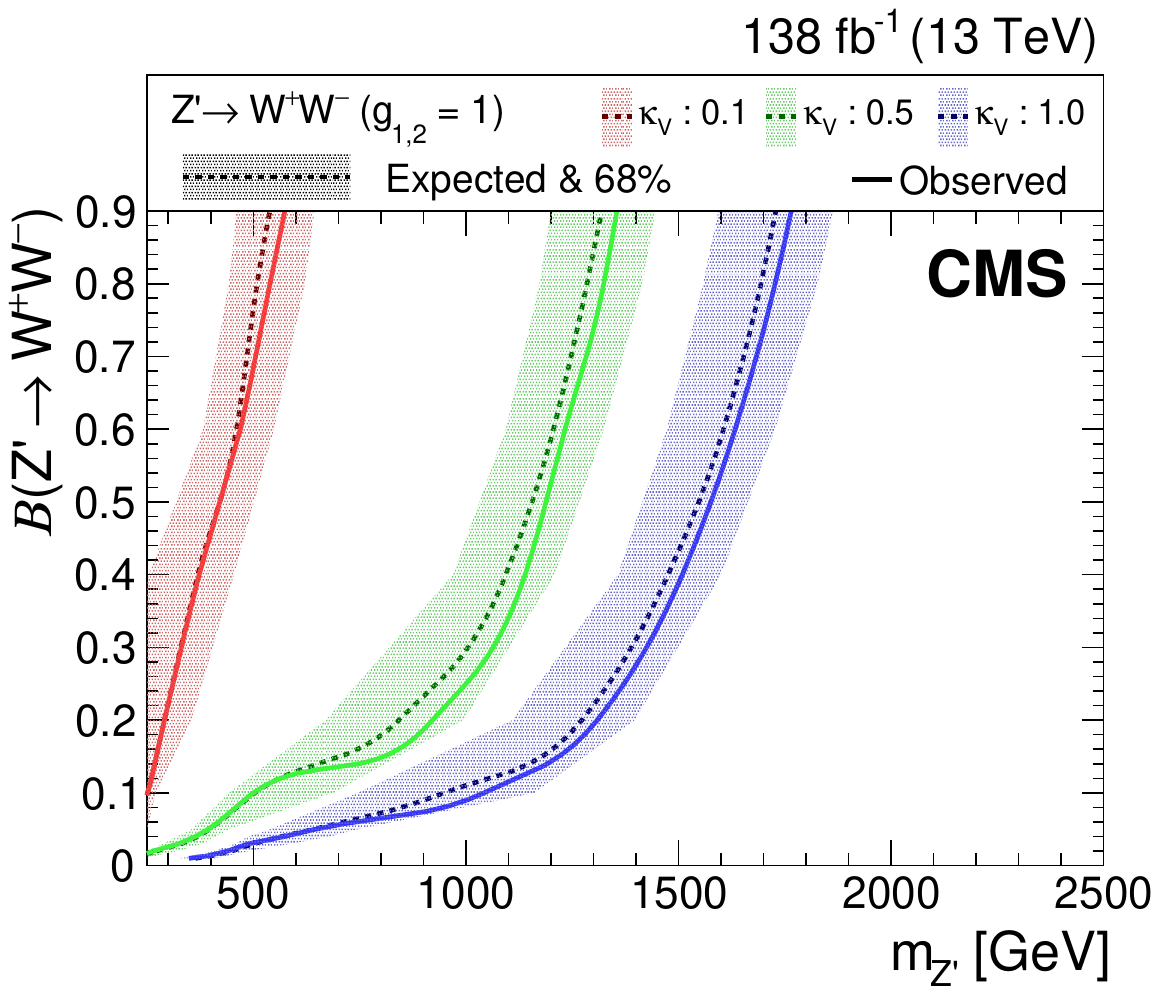}
\caption{Combined 95\% \CL lower limits on \mZpr as a function of the \zp branching fraction to \tautau (upper row) and \WW (lower row) for the $\gl = 0$ (left column) and $\gl = 1$ (right column) scenario. The red, green, and blue curves show the observed limits corresponding to \kV equal to 0.1, 0.5, and 1, respectively.  The dashed curves and shaded bands show the expected limits with their 68\% \CL percentile bands.}
\label{fig:result3}
\end{figure*}

Figure~\ref{fig:result3} shows exclusion bounds for the four signal models described above, as a function of \mZpr and the \Zptautau and \ZpWW branching fractions, assuming different \kV values.

The $\tauh\tauh$ channel contributes most strongly to the \Zptautau exclusion limits, while for \ZpWW it is the \emu channel that contributes most. For SPM1, the data exclude \zp bosons with masses below 1 (2.45) \TeV for a \Zptautau branching fraction of 50\%, assuming $\kV = 0.1$ (1.0). For SPM2, the bounds on \mZpr increase by about 5\% due to an increase in the $\Pp\Pp\to \zp jj$ cross section from pure electroweak non-VBF processes when $\gl = 1$. For SPM3 and SPM4, we exclude \zp with masses below about 1.6\TeV for a 50\% \ZpWW branching fraction, assuming $\kV = 1.0$.
In general, any increment of $\gl$ for constant values of the other couplings lowers the contribution of VBF diagrams relative to the total cross section, thereby decreasing the acceptance of the VBF selections.
Simultaneously, the expected momentum of the \PZpr boson increases, affecting the signal shapes.
For \Zptautau, this leads to weakening limits with increasing $\gl$.
For \ZpWW, because of the mass of the \PW boson, the additional momentum improves the differentiation of the signal shape from that of the background, in total improving the limits with rising $\gl$.
This analysis is sensitive to models that could resolve tensions between SM predictions and precision measurements. Tabulated results are provided in the associated HEPData record~\cite{hepdata}.

In summary, a first search for a heavy neutral spin-1 gauge boson \zp produced via vector boson fusion processes has been performed using data collected by the CMS experiment, corresponding to an integrated luminosity of 138\fbinv. The search considers nonuniversal couplings of \zp bosons to fermions, including scenarios with dominant couplings to the third generation. Two decay channels, \Zptautau and \ZpWW, are investigated. The invariant mass of the dilepton pair plus missing transverse momentum is used to search for the presence of a signal as a broad enhancement above the background expectation. The data are found to be in agreement with standard model predictions. In \zp models with nonuniversal fermion couplings, in particular those with \zp bosons having enhanced couplings to the third generation, the presence of \zp bosons decaying to a \PGt lepton (\PW boson) pair is excluded for \zp masses up to 2.45 (1.60)\TeV, depending on the \zp boson coupling to SM weak bosons and assuming a \Zptautau ($\PWp\PWm$) branching fraction of 50\%. These are the first limits to date on models with nonuniversal fermion couplings.

\begin{acknowledgments}
We congratulate our colleagues in the CERN accelerator departments for the excellent performance of the LHC and thank the technical and administrative staffs at CERN and at other CMS institutes for their contributions to the success of the CMS effort. In addition, we gratefully acknowledge the computing centers and personnel of the Worldwide LHC Computing Grid and other centers for delivering so effectively the computing infrastructure essential to our analyses. Finally, we acknowledge the enduring support for the construction and operation of the LHC, the CMS detector, and the supporting computing infrastructure provided by the following funding agencies: SC (Armenia), BMBWF and FWF (Austria); FNRS and FWO (Belgium); CNPq, CAPES, FAPERJ, FAPERGS, and FAPESP (Brazil); MES and BNSF (Bulgaria); CERN; CAS, MoST, and NSFC (China); MINCIENCIAS (Colombia); MSES and CSF (Croatia); RIF (Cyprus); SENESCYT (Ecuador); ERC PRG, RVTT3 and MoER TK202 (Estonia); Academy of Finland, MEC, and HIP (Finland); CEA and CNRS/IN2P3 (France); SRNSF (Georgia); BMBF, DFG, and HGF (Germany); GSRI (Greece); NKFIH (Hungary); DAE and DST (India); IPM (Iran); SFI (Ireland); INFN (Italy); MSIP and NRF (Republic of Korea); MES (Latvia); LMTLT (Lithuania); MOE and UM (Malaysia); BUAP, CINVESTAV, CONACYT, LNS, SEP, and UASLP-FAI (Mexico); MOS (Montenegro); MBIE (New Zealand); PAEC (Pakistan); MES and NSC (Poland); FCT (Portugal);  MESTD (Serbia); MCIN/AEI and PCTI (Spain); MOSTR (Sri Lanka); Swiss Funding Agencies (Switzerland); MST (Taipei); MHESI and NSTDA (Thailand); TUBITAK and TENMAK (Turkey); NASU (Ukraine); STFC (United Kingdom); DOE and NSF (USA).
\end{acknowledgments}

\bibliography{auto_generated}
\cleardoublepage \appendix\section{The CMS Collaboration \label{app:collab}}\begin{sloppypar}\hyphenpenalty=5000\widowpenalty=500\clubpenalty=5000
\cmsinstitute{Yerevan Physics Institute, Yerevan, Armenia}
{\tolerance=6000
A.~Hayrapetyan, A.~Tumasyan\cmsAuthorMark{1}\cmsorcid{0009-0000-0684-6742}
\par}
\cmsinstitute{Institut f\"{u}r Hochenergiephysik, Vienna, Austria}
{\tolerance=6000
W.~Adam\cmsorcid{0000-0001-9099-4341}, J.W.~Andrejkovic, L.~Benato\cmsorcid{0000-0001-5135-7489}, T.~Bergauer\cmsorcid{0000-0002-5786-0293}, S.~Chatterjee\cmsorcid{0000-0003-2660-0349}, K.~Damanakis\cmsorcid{0000-0001-5389-2872}, M.~Dragicevic\cmsorcid{0000-0003-1967-6783}, P.S.~Hussain\cmsorcid{0000-0002-4825-5278}, M.~Jeitler\cmsAuthorMark{2}\cmsorcid{0000-0002-5141-9560}, N.~Krammer\cmsorcid{0000-0002-0548-0985}, A.~Li\cmsorcid{0000-0002-4547-116X}, D.~Liko\cmsorcid{0000-0002-3380-473X}, I.~Mikulec\cmsorcid{0000-0003-0385-2746}, J.~Schieck\cmsAuthorMark{2}\cmsorcid{0000-0002-1058-8093}, R.~Sch\"{o}fbeck\cmsAuthorMark{2}\cmsorcid{0000-0002-2332-8784}, D.~Schwarz\cmsorcid{0000-0002-3821-7331}, M.~Sonawane\cmsorcid{0000-0003-0510-7010}, W.~Waltenberger\cmsorcid{0000-0002-6215-7228}, C.-E.~Wulz\cmsAuthorMark{2}\cmsorcid{0000-0001-9226-5812}
% ...
\par}
\vskip\cmsinstskip
\dag:~Deceased\\
$^{1}$Also at Yerevan State University, Yerevan, Armenia\\
$^{2}$Also at TU Wien, Vienna, Austria\\
% ...
\end{sloppypar}
\end{document}